\documentstyle[emulateapj,amsfonts,psfig,natbib,amsmath,rotating]{article}
\def\simlt{\mathrel{\hbox{\rlap{\hbox{\lower4pt\hbox{$\sim$}}}\hbox{$<$}}}}
\def\simgt{\mathrel{\hbox{\rlap{\hbox{\lower4pt\hbox{$\sim$}}}\hbox{$>$}}}}

\def\ale{\mathrel{\hbox{\rlap{\hbox{\lower4pt\hbox{$\sim$}}}\hbox{$<$}}}}
\def\age{\mathrel{\hbox{\rlap{\hbox{\lower4pt\hbox{$\sim$}}}\hbox{$>$}}}}

\def\nodata{---}

\def\cit{1}
\def\ociw{2}
\def\pu{3}
\def\hf{4}
\def\uh{5}
\def\srl{6}
\def\toronto{7}
\def\nsf{8}
\def\vla{9}
\def\uva{10}
\def\ucb{11}
\def\ut{12}
\def\llnl{13}
\def\anu{14}

\begin{document}

\title{An {\it HST} Search for Supernovae Accompanying X-ray Flashes}

\author{
A.~M. Soderberg\altaffilmark{\cit},
S.~R. Kulkarni\altaffilmark{\cit}, 
D.~B. Fox\altaffilmark{\cit},
E. Berger\altaffilmark{\ociw,\pu,\hf},
P.~A. Price\altaffilmark{\uh},
S.~B. Cenko\altaffilmark{\srl},
D.~A. Howell\altaffilmark{\toronto},
A. Gal-Yam\altaffilmark{\cit,\hf},
D.~C. Leonard\altaffilmark{\cit,\nsf},
D.~A. Frail\altaffilmark{\vla},
D. Moon\altaffilmark{\srl},
R.~A. Chevalier\altaffilmark{\uva},
M. Hamuy\altaffilmark{\ociw},
K.~C. Hurley\altaffilmark{\ucb},
D. Kelson\altaffilmark{\ociw},
K. Koviak\altaffilmark{\ociw},
W. Krzeminski\altaffilmark{\ociw},
P. Kumar\altaffilmark{\ut},
A. MacFadyen\altaffilmark{\cit},
P.~J. McCarthy\altaffilmark{\ociw},
H.~S. Park\altaffilmark{\llnl},
B.~A. Peterson\altaffilmark{\anu},
M.~M. Phillips\altaffilmark{\ociw},
M. Rauch\altaffilmark{\ociw},
M. Roth\altaffilmark{\ociw},
B.~P. Schmidt\altaffilmark{\anu},
S. Shectman\altaffilmark{\ociw}
}

\altaffiltext{\cit}{Division of Physics, Mathematics and Astronomy,
        105-24, California Institute of Technology, Pasadena, CA
        91125}
\altaffiltext{\ociw}{Observatories of the Carnegie Institution of Washington,
        813 Santa Barbara St., Pasadena, CA 91101}
\altaffiltext{\pu}{Department of Astrophysical Sciences, Princeton University, 
        Princeton, NJ 08544}
\altaffiltext{\hf}{Hubble Fellow}
\altaffiltext{\uh}{University of Hawaii, Institute of Astronomy, 2680 Woodlawn Drive, Honolulu, HI 96822-1897}
\altaffiltext{\srl}{Space Radiation Laboratory 220-47, California Institute of Technology, Pasadena, CA 91125}
\altaffiltext{\toronto}{Department of Astronomy and Astrophysics, University of Toronto, 60 St. George Street, Toronto, ON M5S 3H8, Canada}
\altaffiltext{\nsf}{NSF Astronomy and Astrophysics Postdoctoral Fellow}
\altaffiltext{\vla}{National Radio Astronomy Observatory, Socorro, NM 87801}
\altaffiltext{\uva}{Department of Astronomy, University of Virginia, P.O. Box 3818, Charlottesville, VA 22903-0818}
\altaffiltext{\ucb}{University of California, Berkeley, Space Sciences Laboratory, Berkeley, CA 94720-7450}
\altaffiltext{\ut}{Astronomy Department, University of Texas, Austin, TX 78731}
\altaffiltext{\llnl}{Lawrence Livermore National Laboratory, 7000 East Avenue, Livermore, CA 94550}
\altaffiltext{\anu}{Research School of Astronomy and Astrophysics, The Australian National University, Weston Creek, ACT 2611, Australia}

\begin{abstract}
We present the results from an {\it Hubble Space Telescope/ACS} search
for supernovae associated with X-ray flashes 020903, 040701, 040812
and 040916.  We find strong evidence that XRF\,020903 ($z=0.25$) was
associated with a SN\,1998bw-like supernova and confirm this using
optical spectroscopy at $t\sim 25$ days.  We find no evidence,
however, for SN\,1998bw-like supernovae associated with the other
three events.  In the case of XRF\,040701 ($z=0.21$), we rule out even
a faint supernova similar to SN\,2002ap, using template light-curves
for several local Type Ic supernovae.  For the two cases in which the
redshift is not known, XRFs 040812 and 040916, we derive robust
redshift limits assuming they were accompanied by supernovae
similar to SN\,1998bw and compare these limits with photometric
redshift constraints provided by their host galaxies.  We supplement
this analysis with results for three additional events (XRFs 011030,
020427 and 030723) and discuss the observed diversity of supernovae
associated with X-ray flashes and gamma-ray bursts. We conclude that
XRF-SNe exist, but can be significantly fainter than SN\,1998bw,
possibly consistent with the observed spread in local Type Ibc supernovae.
    
\end{abstract}


\section{Introduction}

Observational evidence for a connection between gamma-ray bursts
(GRBs) and supernovae (SNe) was first established with the discovery
of the highly luminous type Ic SN\,1998bw in spatial and temporal
coincidence with GRB\,980425 \citep{paa+00,gvv+98}.  In the seven
years since this extraordinary event, several possible GRB-SN
associations have been reported based on red ``bumps'' observed in
optical afterglow light-curves (e.g. \citealt{bkd+99}).  Moreover, in
two cases (GRBs 030329 and 031203) there is unambiguous spectroscopic
evidence of high velocity SN features
\citep{hsm+03,smg+03,mgs+04,mtc+04}.  These observations provide
conclusive evidence that at least some gamma-ray bursts are produced
in the explosions of massive stars.
  
In recent years, a new class of high energy transients has been
identified, characterized by an emission spectrum peaking in the X-ray
band, an order of magnitude softer than the peak energies observed for
GRBs \citep{hik+01}.  These so-called X-ray flashes (XRFs) are thought
to be related to GRBs since the two classes share several
observational properties, including prompt emission profiles
(\citealt{slg+04a}, and references therein), broadband afterglows
\citep{bss+04,skb+04a,fsh+04} and star-forming host galaxies at
cosmological distances \citep{bfv+03,lfs+02,lpk+04}.
  
Several hypotheses on the physical connection between XRFs and GRBs
have been proposed.  One popular model posits that XRFs are simply
GRBs viewed away from the jet collimation axis
\citep{yin03,zwh04,grp05}.  In this scenario, the observed prompt emission
is dominated by the mildly relativistic material in
the ``wings'' of the jet, rather than the highly relativistic
($\Gamma\gtrsim 100$) ejecta beamed away from the line-of-sight.
Another popular model suggests that XRFs are produced in a ``dirty
fireball'', where the ejecta carry a more substantial baryonic load
(and hence less relativistic material) than typical GRBs
\citep{zwh04}.  In both scenarios XRFs are expected to be associated
with SNe, whose properties and detectability should not be affected by
the viewing angle or baryonic load.
  
The discovery of a SN in association with an XRF would therefore
conclusively associate XRFs with the death of massive stars and hence
GRBs.  Motivated thus, we undertook a systematic search for SNe
associated with XRFs using the {\it Hubble Space Telescope} ({\it
HST}).  As part of our XRF-SN analysis, we synthesized supernova
light-curves at various redshifts utilizing as templates the
well-sampled optical light-curves of several local SNe
(\S\ref{sec:templates}).  Comparison of the synthesized SNe with
our {\it HST} observations enabled us to study the diversity of XRF
associated SNe.  Details on the individual {\it HST} targets and
observations follow in \S\ref{sec:targets}.  By including results from
other XRF-SN searches, we compile an extended sample of seven events
(XRFs 011030, 020427, 020903, 030723, 040701, 040812 and 040916), and
present a global summary of XRF-SN detection limits in
\S\ref{sec:summary}.  A discussion on the observed spread in the peak
optical luminosities of GRB- and XRF-associated SNe follows as
\S\ref{sec:conclusions}.

\section{Supernova Light-Curve Synthesis}
\label{sec:templates}

In modeling the XRF-associated SNe, we adopted optical data for the
local SNe 1994I, 1998bw and 2002ap as templates.  These three SNe were
selected based on their well-sampled optical light-curves which
represent an overall spread in the observed properties of Type Ibc
supernovae.  To produce synthesized light-curves for each of these
template SNe, we compiled optical $UBVRI$ observations from the
literature and smoothed the extinction-corrected (foreground plus host
galaxy) light-curves.  We then redshifted the light-curves by
interpolating over the photometric spectrum and stretching the arrival
time of the photons by a factor of $(1+z)$.  Since observed spectra of
local ($d\lesssim 100$ Mpc) Type Ibc SNe show a steep drop-off in flux
blue-ward of $\sim 4000$ \AA\ due to heavy line-blanketing and
since good-quality UV data are currently not available below $3000$
\AA\, we do not attempt to extrapolate the rest-frame spectra
blue-ward of the rest-frame $U$-band observations.  This limits the
synthesized light-curves to $z\le 0.80$ and $z\le 1.20$ for the
observed $R$- and $I$-bands, respectively.  Below we discuss the
compiled optical datasets for each of the template SNe.

\subsection{SN\,1998bw}
\label{sec:templates_SN1998bw} 
The well-sampled $UBVRI$ light-curves for SN\,1998bw were taken from
\citet{gvv+98} and \citet{ms99} and corrected for Galactic extinction
($A_V=0.19$; \citealt{sfd+98}).  We assume negligible host galaxy
extinction, consistent with the spectroscopic analysis \citet{pcd+01}.
The broadband optical dataset spans a timescale from $t\approx 0.7$ to
417 days.  Here, the explosion time is set by the {\it Beppo-SAX}
detection of GRB\,980425 on 1998 April 25.91 UT \citep{paa+00}.  In
calculating optical luminosities for SN\,1998bw, we assume a distance,
$d_L \approx 36.1$ Mpc ($H_0=71~\rm km/s/Mpc$, $\Omega_{M}=0.27$,
$\Omega_{\Lambda}=0.73$), based on the observed redshift to the host
galaxy, ESO 184-G82 \citep{gvp+98}.

\subsection{SN\,1994I}
\label{sec:template_SN1994I}
\citet{rvh+96} provide a large compilation of multi-color light-curves
for SN\,1994I.  We adopt a large host galaxy extinction of $A_V=1.4$
and negligible Galactic extinction as derived through the spectroscopic
analysis \citep{rvh+96}.  Using an explosion date of 1994 March 30
UT from radio light-curve modeling \citep{ssw+04}, the $BVRI$ data
span from $t\approx 1.4$ to 130 days, while the $U$-band data extend
only as far as $t\approx 47.4$ days. In an effort to extend the
$U$-band light-curve, we scale the late-time linear decay of the
$B$-band light-curve to match the last epoch of $U$-band observations
and assume the $U-B$ color is constant thereafter.  We note that this
scaling introduces a source of uncertainty in our late-time [$t\gtrsim
  47(1+z)$] high$-z$ synthesized light-curves of SN\,1994I.  In
calculating optical luminosities, we adopt a distance of $d_L\approx
8.5$ Mpc for host galaxy, M51, as given by \citet{rvh+96}.

\subsection{SN\,2002ap}
\label{sec:templates_SN2002ap}
$UBVRI$ light-curves were taken from \citet{fps+03a} and scaled to an
explosion date of 2002 January 28.9 UT \citep{mdm+02}.  We adopt the
spectroscopic derived total extinction (foreground plus host galaxy)
of $A_V=0.26$ \citep{fps+03a}. Data span $t\approx 1.3$ to 317.3 days
after the explosion in $BVRI$ filters while the $U$-band data extend
only to $t\sim 35.2$ days.  In a manner similar to that for SN\,1994I,
we extend the $U$-band light-curve by scaling the late-time $B$-band
data and note that this introduces uncertainty in the synthesized
light-curves at high$-z$.  We assume the distance to the host galaxy, M74,
is $d_L \approx 7.3$ Mpc \citep{skt96,sd96}.

We emphasize the striking differences between the three SN
light-curves when the extinction-corrected rest-frame $V$-band
light-curves are compared.  With regard to the luminosity at peak time,
SN\,1998bw is a factor of $\sim 2.2$ more luminous than SN\,1994I and
$\sim 6$ times more luminous than SN\,2002ap.  Moreover, the time of
$V$-band peak vary by a factor of two: while SN\,1998bw peaks at
$t\sim 16$ days, SN\,1994I and SN\,2002ap both peak at just $t \sim 9$
days. Such early peak times present a challenge for GRB-SN searches, since
the optical afterglow typically dominates on these timescales.

\section{Hubble Space Telescope XRF-SN Search}
\label{sec:targets}

Since the activation of our Cycle-13 {\it HST} program to
study the supernovae associated with X-ray flashes and gamma-ray
bursts (GO-10135; PI: Kulkarni), three XRFs have been discovered and
localized by their afterglow emission: XRFs 040701, 040812 and 040916.
In an effort to study the SNe possibly associated with these XRFs, we
observed each of these objects with {\it HST} at late-time, when
an associated supernova is most likely to dominate the optical
emission.  To supplement our sample of XRF observations,
we investigated archival {\it HST} images of XRF\,020903 (GO-9405;
PI: Fruchter). We describe our data analysis techniques below.

Using the {\it Wide-Field Camera} ({\it WFC}) of the {\it Advanced
  Camera for Surveys} ({\it ACS}) on-board {\it HST}, we imaged the
  fields of XRFs 040701, 040812 and 040916.  For each target we
  undertook observations at two epochs, $t\sim 30$ and $\sim 60$ days,
  in order to search for optical emission associated with an
  underlying supernova.  Each epoch consisted of two orbits during
  which we imaged the field in two filters, F625W and F775W,
  corresponding to SDSS $r'$- and $i'$- bands, respectively.

We retrieved archival images of XRF\,020903 from the {\it HST}
  archive\footnote{http://archive.stsci.edu/hst/search.php}.  Similar
  to the other bursts in our sample, the XRF\,020903 data were
  obtained with {\it HST/ACS} using {\it WFC}.  We analyze the images
  from two epochs at $t\sim 91$ and $\sim 300$ days to search for the
  signature of an associated supernova.  These data were taken in the
  broad $V$-band filter, F606W.

The {\it HST} data were processed using the {\tt multidrizzle} routine within
the {\tt stsdas} package of IRAF \citep{fh02}.  Images were drizzled
using {\tt pixfrac}=0.8 and {\tt pixscale}=1.0 resulting in a final
pixel scale of 0.05 arcsec/pixel.  Drizzled images were then
registered to the first epoch using the {\tt xregister} package within
{\it IRAF}.

To search for source variability, we used the {\it ISIS} subtraction
routine by \citet{a00} which accounts for temporal variations in the
stellar PSF.  Residual images (Epoch 1 $-$ Epoch 2) were examined for
positive sources positionally coincident with the afterglow error
circle.  To test our efficiency at recovering transient sources, false
stars with a range of magnitudes were inserted into the first epoch
images using {\it IRAF} task {\tt mkobject}.  An examination of the
false stellar residuals provided an estimate of the magnitude limit
($3\sigma$) to which we could reliably recover transients.

Photometry was performed on the residual sources within a 0.5 arcsec
aperture.  We converted the photometric measurements to infinite
aperture and calculated the corresponding AB magnitudes within the native
{\it HST} filters using the aperture corrections and zero-points
provided by \citet{sjb+05}.  For comparison with ground-based data, we
also converted the photometric measurements to Johnson 
$R$- and $I$-band (Vega) magnitudes using the transformation coefficients
derived by \citet{sjb+05} and assuming a flat $F_{\nu}$ source spectrum.

In the following sections we summarize the afterglow properties for
each of the targets and the photometry derived from our {\it HST}
SN search.  A log of the {\it HST} observations for the four XRFs follows in
Table~\ref{tab:hst}.

\subsection{XRF\,020903} 
\label{sec:xrf020903}

\subsubsection{Prompt Emission and Afterglow Properties}
\label{sec:ag_020903}

XRF\,020903 was detected by the Wide-Field X-ray Monitor ({\it WXM})
on-board the High Energy Transient Explorer ({\it HETE-2}) satellite
on 2002 September 3.421 UT.  With a spectral energy distribution
peaking below 5 keV, XRF\,020903 is the softest burst detected
during the lifetime of the instrument \citep{slg+04b}.  Despite the
large $4\times 31$ arcmin localization region, an optical afterglow
was discovered \citep{skb+04a} at $\alpha=22^{\rm h} 48^{\rm m}
42^{\rm s}.34$, $\delta=-20^{\circ} 46' 09''.3$ (J2000).  At $t\approx
0.9$ days, the afterglow had $R\approx 19.5$ mag and continued to fade
as $F_{\rm opt} \propto t^{-1.1}$ until $t\sim 30$ days when the decay
flattened to a plateau.  Optical spectroscopy showed the transient
source to be associated with a galaxy complex at $z=0.251$ \citep{spf+02}.

\subsubsection{{\it HST} Observations}
\label{sec:hst_020903}

XRF\,020903 was observed using {\it HST/ACS} on 2002 December 3.79 and
2003 June 30.65 UT, corresponding to $t\sim 91$ and 300 days after the
burst.  Imaging was carried out in the broad F606W filter for total
exposure times of 1840 sec (Epoch 1) and 1920 sec (Epoch 2).
Relative astrometry was performed using an early-time ($t\sim 0.9$ day) image
from the Palomar 200-inch telescope \citep{skb+04a}.  Using 42
unsaturated, unconfused stars in common between the two images, we
registered the {\it HST} data with a systematic uncertainty
of 0.12 arcsec ($2\sigma$).

\begin{figure}
\vspace{-1.5cm}
\plotone{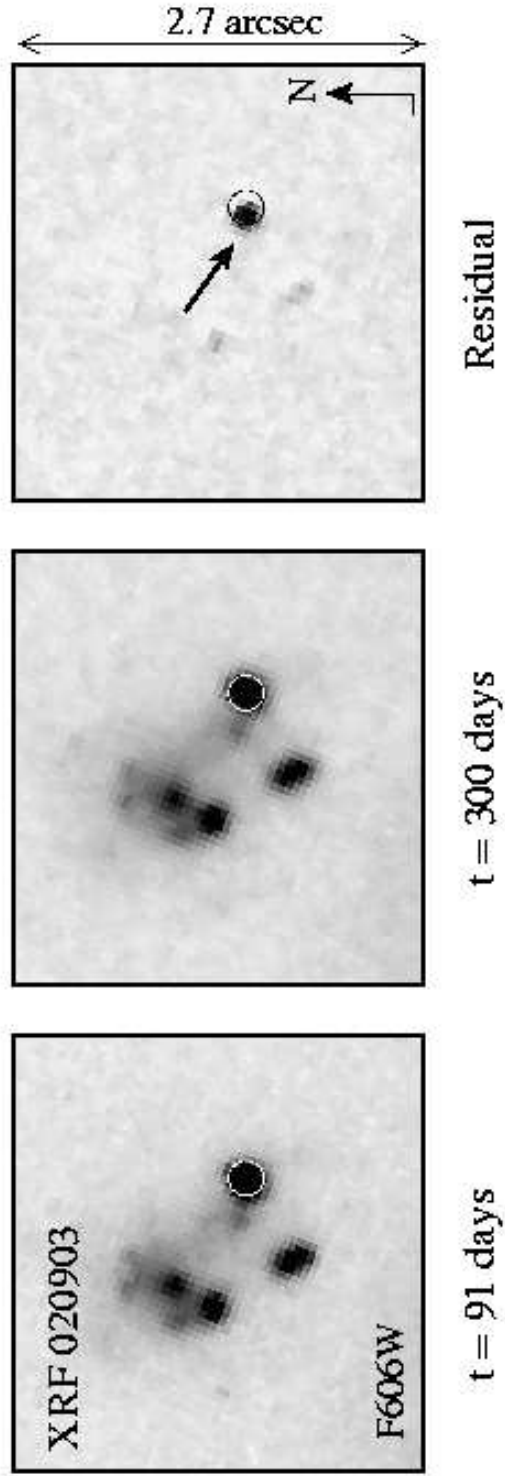}
\vspace{0cm}
\caption{Three-panel frame showing {\it HST/ACS} imaging for
  XRF\,020903 at $t\sim 91$ days (Epoch 1) and $t\sim 300$ days (Epoch
  2) in the F606W filter. By subtracting the Epoch 2 image from Epoch
  1, we produced the residual image, above.  We apply
  the same stretch to all frames. As clearly shown in this residual
  image, a transient source is detected coincident with the $0.12$
  arcsec (2$\sigma$) optical afterglow position (circle), lying on
  the southwest knot of the host galaxy complex.
\label{fig:XRF020903_HST}}
\end{figure}

The {\it HST} images reveal that the afterglow localization circle
coincides with the southwest knot of the host galaxy complex
(Figure~\ref{fig:XRF020903_HST}).  Through image subtraction, we find
a positive residual coincident with the optical afterglow position at
$\alpha=22^{\rm h} 48^{\rm m} 42.293^{\rm s}$, $\delta=-20^{\circ} 46'
08''.47$ (J2000).  Photometry of the residual source gives
F606W$\approx 24.53\pm 0.03$ mag ($R\approx 24.32\pm 0.03$ mag).
Correcting for Galactic extinction ($A_R=0.09$; \citealt{sfd+98}), the
true magnitude of the source is $R\approx 24.23 \pm 0.03$ mag.  To
estimate our photometric uncertainty, we placed random apertures near
other galaxy residuals and calculated the standard deviation of the
resulting values.

\begin{figure}
\vspace{0cm}
\plotone{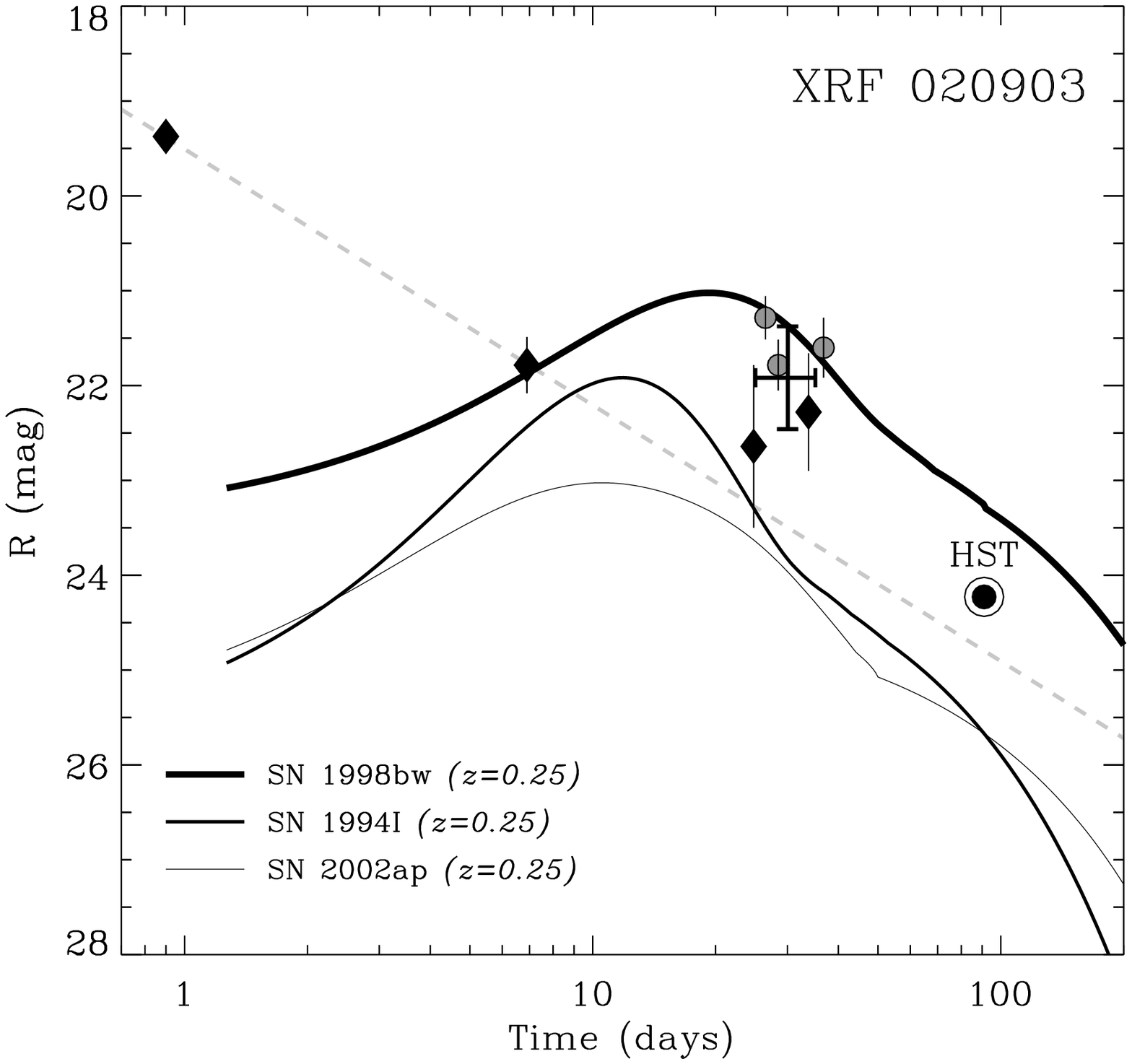}
\vspace{0cm}
\caption{Constraints on a supernova associated with XRF\,020903.
  Extinction corrected ground-based $R$-band observations of the
  optical afterglow have been compiled from \citet{skb+04a} and the
  GCNs \citep{cmg+02,ghp+02}, and are shown as diamonds and grey circles,
  respectively.  The temporal decay of the (host galaxy subtracted)
  optical afterglow is described by $t^{-1.1}$ (dashed grey line) at
  early time followed by a plateau phase at $t\sim 30$ days.  The
  weighted mean of the points between $t\sim 20-40$ days (black cross)
  indicates that at peak the supernova was $0.6\pm 0.5$ magnitudes
  fainter than the synthesized SN\,1998bw light-curve (thick line).
  At $t\sim 91$ days, however, the {\it HST} transient (encircled dot)
  is $\sim 1$ magnitude fainter than the synthesized curve, implying
  the supernova faded faster than SN\,1998bw.  SN\,1994I- and
  2002ap-like supernovae (medium and thin lines, respectively) would
  be significantly fainter than the {\it HST} residual.
\label{fig:XRF020903_SN_curve}}
\end{figure}

Figure~\ref{fig:XRF020903_SN_curve} shows the extinction-corrected
{\it HST} photometry along with ground-based $R$-band data for the
optical afterglow associated with XRF\,020903.  Ground-based data have
been compiled from the GCNs \citep{cmg+02,ghp+02} as well as from
Table 1 of \citet{skb+04a}.  We have numerically subtracted the host
galaxy contribution from the ground-based data, assuming an
extinction-corrected host galaxy brightness of $R\approx 20.90$ mag
based on late-time observations \citep{ghp+02,lfs+02}.  From the
compiled $R$-band afterglow light-curve, it is evident that the
temporal decay flattens significantly around $t\sim 30$ days and
subsequently steepens toward the {\it HST} measurement at $t\sim 91$
days.  This flattening (or plateau phase)
occurs on the same timescale that a SN\,1998bw-like supernova at
$z=0.25$ would reach maximum light (see also \citealt{bfr+04}).

\subsubsection{Associated Supernova}
\label{sec:sn_020903}

Over-plotted in Figure~\ref{fig:XRF020903_SN_curve} are the
synthesized light-curves of SNe 1998bw, 1994I and 2002ap at a redshift
of $z=0.25$.  It is clear that an associated SN\,1998bw-like supernova
would be $\sim 1$ magnitude brighter than the {\it HST} observation at
$t\sim 91$ days, while SN\,1994I and SN\,2002ap-like light-curves are
each fainter by 1.4 magnitudes.  By taking the weighted average of the
ground-based data between $t\sim 20 - 40$ days, we predict that the
supernova was $0.6\pm 0.5$ magnitudes fainter than SN\,1998bw at
maximum light.  We note that this uncertainty is dominated by
``aperture effects'' (see \citealt{pks+03}) which cause variable
contribution from the host galaxy complex in different epochs.

\begin{figure}
\plotone{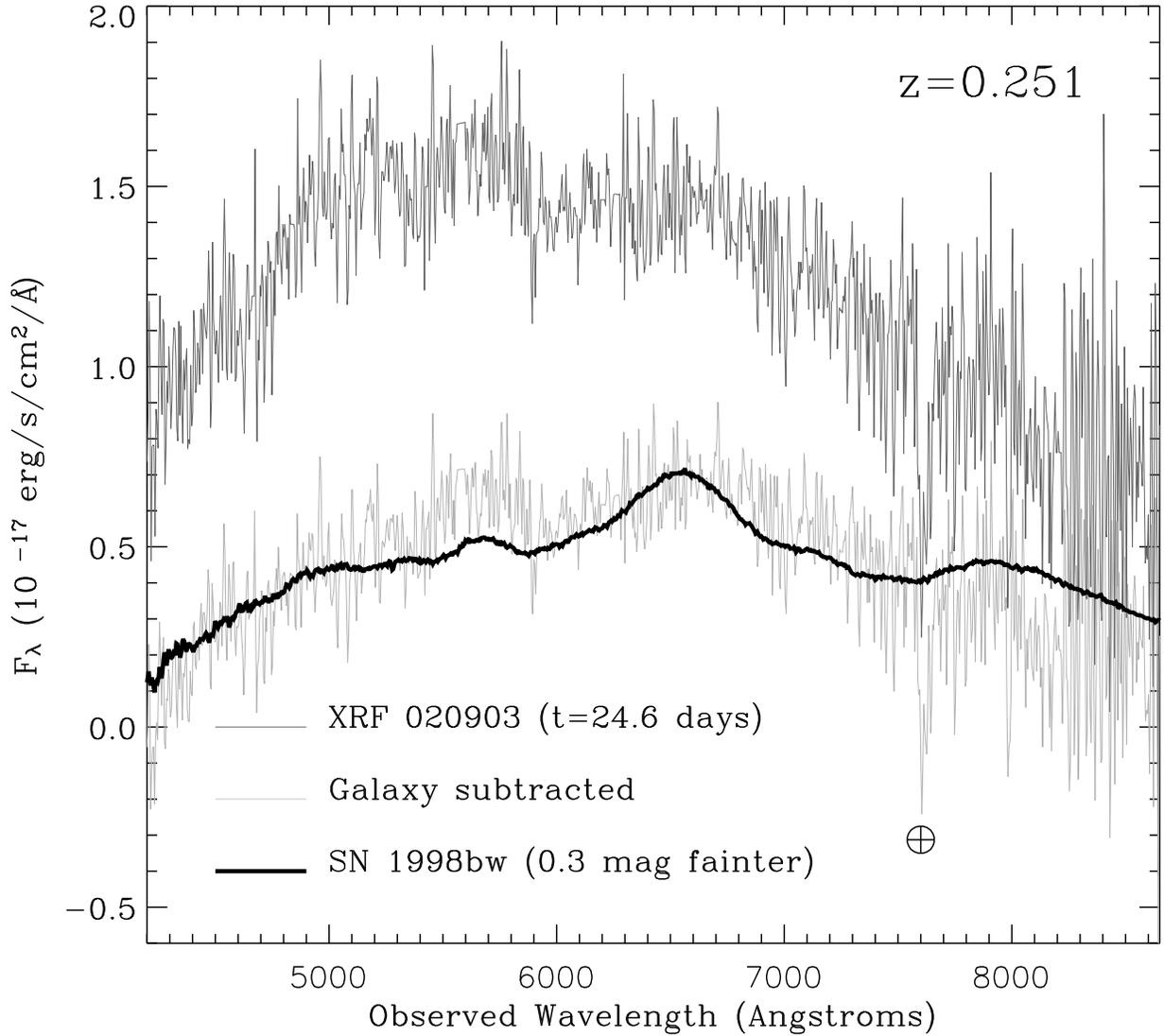}
\caption{Magellan/LDSS2 spectrum of XRF\,020903, taken at $t\approx
24.6$ days.  Host galaxy emission lines were sigma-clipped (top) and
we remove the host galaxy contribution using a starburst template
(bottom; light grey).  The galaxy subtracted spectrum has broad SN features, in
clear resemblance to SN\,1998bw at $t\sim 20$ days (rest-frame),
redshifted to $z=0.25$ and dimmed by $0.3$ magnitudes (bottom; black).
The telluric band is marked with an encircled cross. 
\label{fig:XRF020903_SN1998bw}}
\end{figure}

We obtained optical spectroscopy of the transient source using the
Magellan 6.5-meter telescope equipped with the Low Dispersion Survey
Spectrograph (LDSS2) on 2002 September 28.06 UT ($t\approx 24.6$
days), during the observed plateau phase.  The data were reduced and
calibrated using standard techniques.  The spectrum is characterized
by a faint continuum dominated by narrow, bright emission lines
typical of star-forming galaxies (Figure~\ref{fig:XRF020903_SN1998bw};
\citealt{spf+02,cf02}).

To search for high velocity SN features within the observed spectrum,
we utilized the supernova classification techniques of \citet{hsp+05},
designed for identification of SNe in the presence of host galaxy
contamination.  Host galaxy light must be subtracted from the observed
spectrum to reveal the SN flux.  We fit a range of starburst host
galaxy templates from \citet{kbc+93}, consistent with the continuum
shape and narrow lines in the observed spectrum.  After sigma-clipping
the narrow emission lines and subtracting the best-fit galaxy template
(model SB1), the residual spectrum shows broad features resembling
those of SN\,1998bw near maximum light.
Figure~\ref{fig:XRF020903_SN1998bw} presents a comparison of our
galaxy subtracted spectrum with SN\,1998bw at $t\approx 20.5$ days
(rest-frame; \citealt{pcd+01}), redshifted to $z=0.25$ and dimmed by
$\sim 0.3$ magnitudes.  The resemblance is striking.

Taken together, the spectroscopic and photometric data strongly
suggest that XRF\,020903 was associated with a supernova that is $\sim
0.6\pm 0.5$ magnitudes fainter than SN\,1998bw at maximum
light. Moreover, the SN light-curve fades faster than SN\,1998bw,
falling $\sim 1$ mag below the synthesized SN\,1998bw light-curve at
late-time.  A dimmer, faster fading supernova was also interpreted for
SN\,2003dh/GRB\,030329 (\citealt{log+04,dtm+05};
c.f. \citealt{mgs+04}) and is consistent with the luminosity-stretch
relation for GRB-SNe \citep{bkp+02,sgn+05}.

\subsection{XRF\,040701}
\label{sec:xrf040701}

\subsubsection{Prompt Emission and Afterglow Properties}
\label{sec:ag_040701}

XRF\,040701 was localized on 2004 July 1.542 UT by the {\it HETE-2} WXM
to an 8 arcmin radius error circle centered at $\alpha=20^{\rm h} 47^{\rm m}
46.3^{\rm s}$, $\delta=-40^{\circ} 14' 13''$ (J2000; \citealt{bra+04}).  

We observed the error circle with the {\it Chandra} X-ray Observatory
(CXO) using the AXAF CCD Imaging Spectrometer (ACIS) for 22.3 ksec
beginning at 2004 July 9.32 UT ($t \approx 7.9$ days) and 20.4 ksec on
2004 July 18.05 ($t \approx 16.6$ days).  Comparison of the two epochs
revealed the most variable source to be at position $\alpha=20^{\rm h}
48^{\rm m} 16^{\rm s}.097$, $\delta=-40^{\circ} 11' 08''.83$ (J2000),
with an uncertainty of $0.5$ arcsec in each coordinate ($2\sigma$;
\citealt{f04b}) which we interpret as the X-ray afterglow.  Assuming
Galactic absorption, the X-ray flux of the source was $7.75\times
10^{-14}$ and $4.06 \times 10^{-14}~\rm erg~cm^{-2}~s^{-1}$ ($2-10$
keV), in the first and second epochs, respectively.  This implies a
temporal decay, $F_{X}\sim t^{\alpha}$ with $\alpha=-1.15$, between
the two observations, comparable to the typical observed values of GRB
X-ray afterglows, $\alpha\sim -1.1$ (\citealt{bkf03}, and references
therein).

Inspection of Digital Sky Survey (DSS) images revealed that the X-ray
afterglow is associated with a resolved galaxy complex whose redshift
we determined to be $z=0.2146$ \citep{kkb+04}.  At this relatively low
redshift, the X-ray afterglow is sensitive to absorption within the
host galaxy.  We fit an absorbed power-law model to the afterglow
spectrum where the column density, $N_H$, was a combination of
foreground and host galaxy extinction.  We find that the column
density within the host galaxy must be $N_H \lesssim 5\times
10^{21}~\rm cm^{-2}$ (90$\%$ confidence) in order to reproduce the
observed low energy ($< 1$ keV) X-ray photons.  Utilizing the $N_H$ to
$A_V$ conversion of \citet{ps95}, this limit corresponds to a
rest-frame host galaxy extinction of $A_{V,\rm host}\lesssim 2.8$ mag.

Despite deep searches, no optical afterglow candidate was discovered
through ground-based monitoring of the {\it Chandra} error circle
\citep{uts+04,bgf+04,pfj+04}.  This non-detection could be the result
of the large host-galaxy extinction, consistent with the observed
X-ray afterglow absorption.  

\subsubsection{{\it HST} observations}
\label{sec:hst_040701}

{\it HST/ACS} imaging was carried out on 2004 August 9.66 and 30.52 UT
($t \approx 39.1$ and 60.0 days after the burst).  Each epoch had a
total exposure time of 1840 and 1920 sec in the F625W and F775W
filters, respectively.  We astrometrically tied the {\it HST} and
{\it Chandra} images by first registering the X-ray source list
to our $I$-band images from the Las Campanas Observatory (LCO) 40-inch
telescope \citep{bgf+04} using three sources in common.  We then tied
the LCO images to those from {\it HST} resulting in a final
positional uncertainty of 1.06 arcsec ($2\sigma$).  

\begin{figure}
\vspace{-2cm}
\plotone{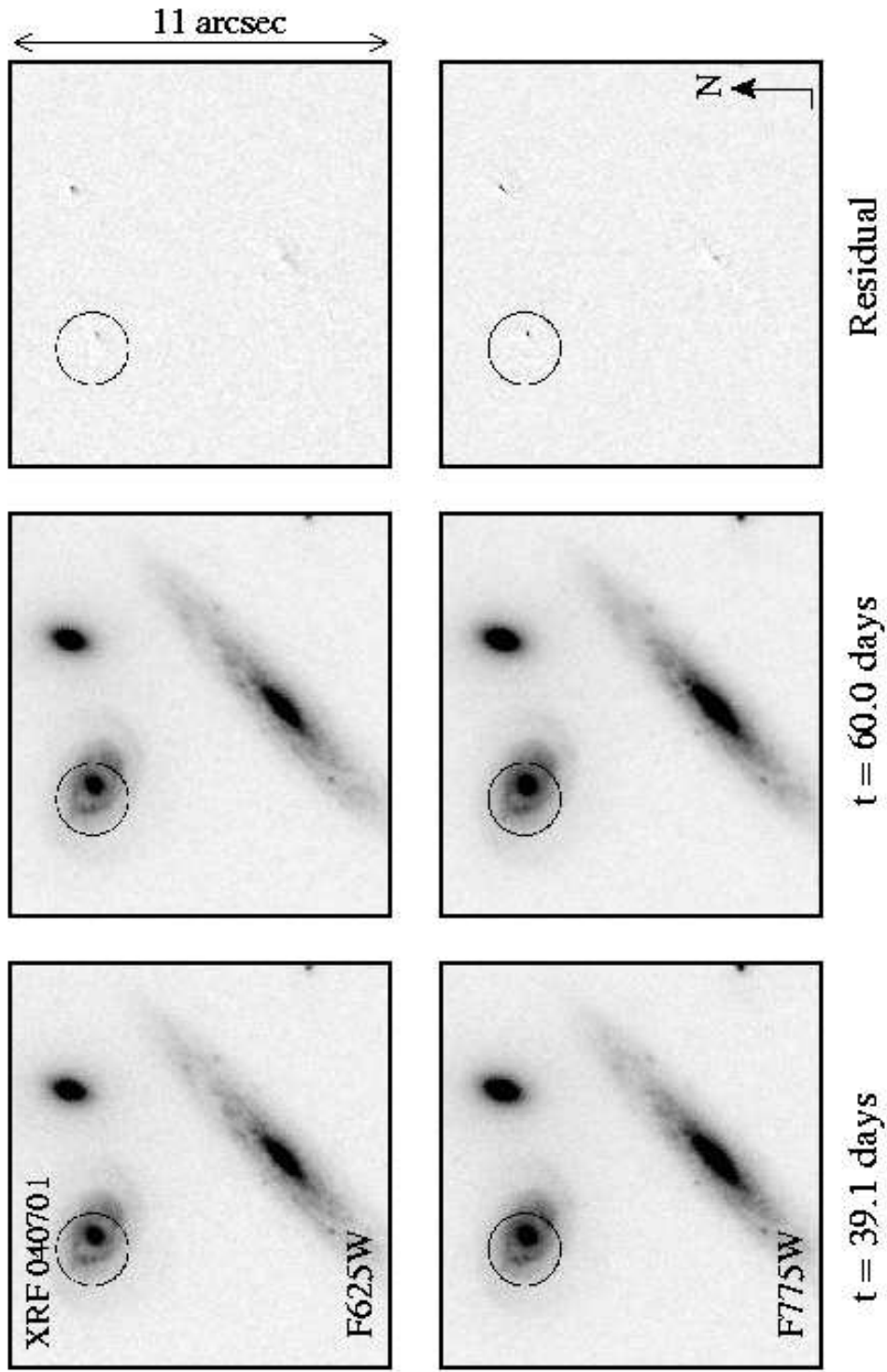}
\vspace{0cm}
\caption{Six-panel frame showing {\it HST/ACS} imaging for XRF\,040701
  at $t\sim 39.1$ days (Epoch 1) and $t\sim 60.0$ days (Epoch 2) in
  the F625W and F775W filters. By subtracting Epoch 2 images from
  Epoch 1, we produced the residual images, above.  We apply the same
  stretch to all frames.  As shown in these residual images, there is
  no evidence of an optical transient within the $\it Chandra$ X-ray
  afterglow position ($2\sigma$; circle) down to our {\it HST/ACS}
  detection limits.
\label{fig:XRF040701_HST}}
\end{figure}

Our {\it HST} observations reveal that the afterglow error circle
coincides with the northeast galaxy of the host complex.  Inspection
of the images reveals that there are no transient sources within this
localization region.  Figure~\ref{fig:XRF040701_HST} shows the images
from both epochs in addition to the residual images produced from the
subtraction routine.  We found that the false stellar residuals
recovered at our detection threshold correspond to limits of F625W $>
27.8$ and F775W $ > 26.8$ mag ($R > 27.6$ and $I > 26.4$ mag).
Correcting for Galactic extinction ($A_R=0.13$; \citealt{sfd+98}), the
actual limits are $R > 27.5$ and $I > 26.3$ mag.  We note that the
slightly elevated flux of the bipolar galaxy residual are consistent
with these limits.

\begin{figure}
\vspace{-1.5cm}
\plotone{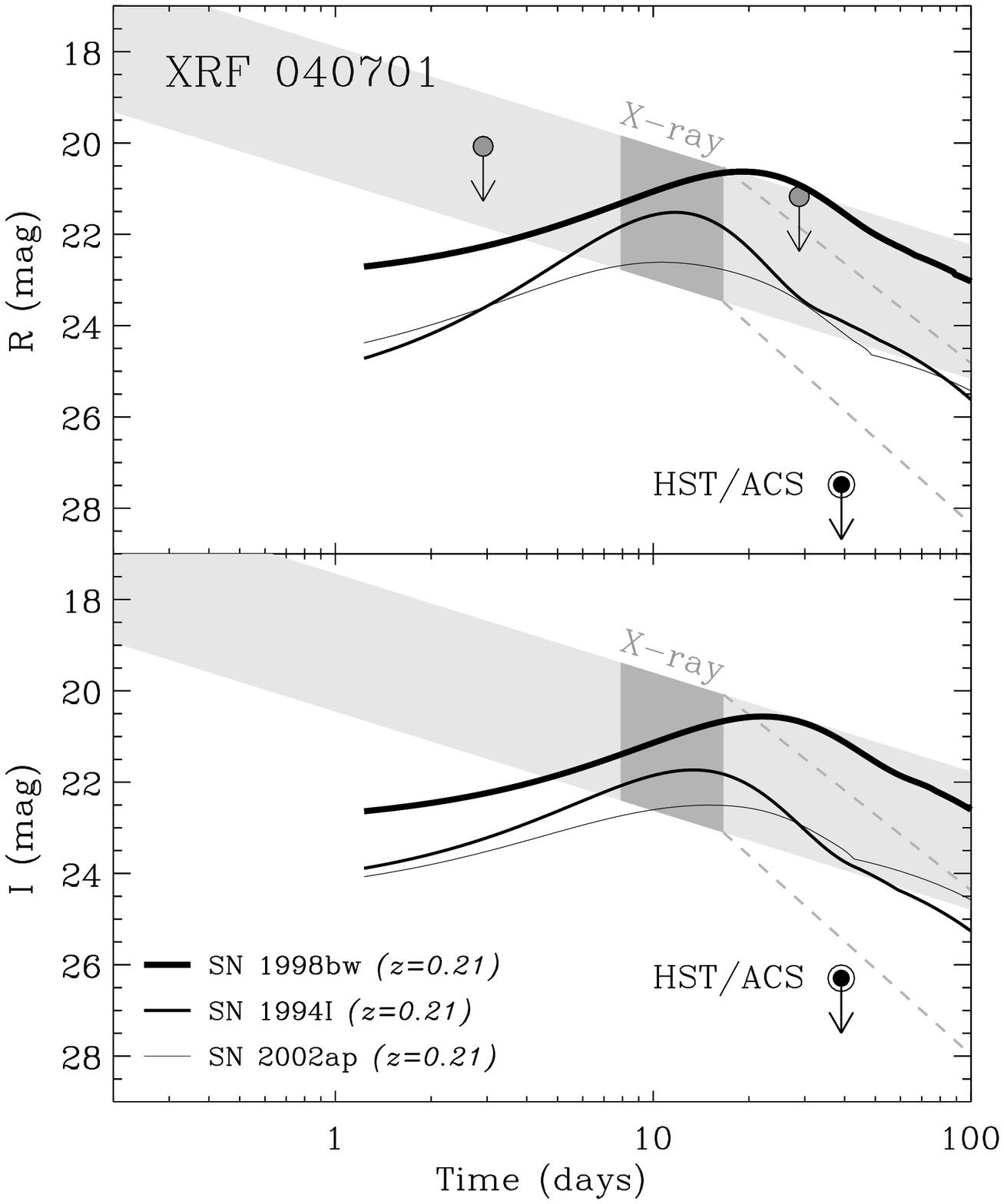}
\vspace{-1.cm}
\caption{Constraints on an optical transient associated with
  XRF\,040701.  Ground-based optical limits (grey circles) have been
  compiled from the GCNs \citep{uts+04,pfj+04}, corrected for Galactic
  extinction and are plotted for the $R$- (upper panel) and $I$-band
  (lower panel).  By extrapolating the observed X-ray afterglow to the
  optical bands, we predict a range of magnitudes for the optical
  afterglow on these same timescales (dark shaded polygon).  Assuming
  the temporal decay can be extrapolated outside the X-ray observation
  window produces the light shaded bands.  Dashed lines represent the
  extreme case where a jet break occurs at the second {\it Chandra}
  epoch.  Simulated SN light-curves for SNe 1998bw (thick line), 1994I
  (medium line) and 2002ap (thin line), redshifted to $z=0.21$, are
  over-plotted. Image subtraction of our two {\it HST/ACS} epochs in
  F625W and F775W, provide deep limits on an associated SN at $t\sim
  32.1$ days (encircled dots).
\label{fig:XRF040701_SN_curve}}
\end{figure}

Figure~\ref{fig:XRF040701_SN_curve} shows the {\it HST} limits along
with early-time data compiled from the GCNs \citep{uts+04,pfj+04},
corrected for Galactic extinction. To estimate the flux of the optical
afterglow, we extrapolated the observed {\it Chandra} data to the $R$-
and $I$-bands.  Following \citet{spn98} for the case of a constant
denisty medium, the observed X-ray afterglow decay ($F_{X}\propto
t^{-1.15}$) implies that the electron energy index ($dN/dE\propto
E^{-p}$) is $p\approx 2.5$ in the case where the synchrotron cooling
frequency, $\nu_c$, is above the X-ray band, and $p\approx 2.2$, if it
is below.  We therefore extrapolate the observed X-ray flux by
adopting a spectral index ($F_{\nu}\propto \nu^{\beta}$) with
$\beta=-(p-1)/2 \approx -0.75$ in the case of $\nu_{\rm opt} < \nu_X <
\nu_c$ and $\beta=-p/2 \approx -1.1$ for $\nu_c < \nu_{\rm opt} <
\nu_X$.  These two scenarios bracket the whole range of optical flux
values implied by the possible location of the $\nu_c$ at the time of
the {\it CXO} observations.

By extrapolating the predicted optical decay to the first {\it HST}
epoch, we conclude that the afterglow should have been $\gtrsim 3.0$
magnitudes brighter than our {\it HST} detection limit.  Even in the
most extreme scenario, a jet break occurred coincident with the second
{\it Chandra} epoch, forcing the temporal decay to steepen to
$F_{\nu}\propto t^{-p}$ \citep{sph99}.  Still, the afterglow would
have been $\gtrsim 1.7$ magnitudes brighter than the {\it HST} limit.
Under the assumption that the observed X-ray flux was dominated by
synchrotron emission, these limits imply that there is significant
extinction from the host galaxy.  However, in a scenario where the
X-ray emission is dominated by other processes (e.g. inverse Compton),
this extrapolation over-predicts the brightness of the optical
afterglow, thereby reducing the implied host galaxy extinction.

\subsubsection{Supernova Limits}
\label{sec:sn_040701}

Over-plotted in Figure~\ref{fig:XRF040701_SN_curve} are synthesized
light-curves for SNe 1998bw, 1994I and 2002ap at $z=0.21$.  From the
figure, it is clear that a SN\,1998bw-like supernova would have been
$\sim 6$ magnitudes brighter than our {\it HST} limits.  A faint
supernova similar to SN\,2002ap would still be $\sim 3.4$ magnitudes
above our detection threshold.  Our constraints on the column density
imply that host galaxy extinction cannot account for this difference;
even in an extreme scenario, given by $A_{V,\rm host}\approx 2.8$ mag,
our limits are still $\sim 3.2$ magnitudes fainter than SN\,1998bw.
We conclude that an XRF\,040701-associated supernova must be $\sim 3$
mag fainter than SN\,1998bw, making it significantly fainter than all
GRB-SNe known to date.

\subsection{XRF\,040812}
\label{sec:xrf040812}

\subsubsection{Prompt Emission and Afterglow Properties}
\label{sec:ag_040812}

XRF\,040812 was discovered on 2004 August 12.251 UT by the Imager on
Board the Integral Satellite (IBIS).  Preliminary analysis indicated a
spectrum that was X-ray rich.  The event was localized to a 2-arcmin
radius circle centered at $\alpha=16^{\rm h} 26^{\rm m} 05^{\rm s}$,
$\delta=-44^{\circ} 42' 32''$ (J2000; \citealt{gmm+04}).

\citet{pkm+04a} observed the field of XRF\,040812 with {\it
  Chandra}/ACIS beginning on 2004 August 17.30 UT ($t\approx 5$ days)
and on 2004 August 22.41 UT ($t \approx 10$ days) for 10 ksec each.
Comparison of the two epochs revealed a variable source at position
$\alpha=16^{\rm h} 26^{\rm m} 2^{\rm s}.25$, $\delta=-44^{\circ} 43'
49''.4$ (J2000) which faded as $F_{X} \propto t^{-1.4}$ between the
observations \citep{pkm+04b}. The unabsorbed flux in the first and
second epochs was $2.53\times 10^{-13}$ and $9.30\times 10^{-14}~\rm
erg~cm^{-2}~s^{-1}$ (0.5--10 keV), respectively \citep{cm04a,cm04b}.

Due to high Galactic extinction in the direction of the burst
($A_R=3.6$ mag; \citealt{sfd+98}) and the presence of an extremely
bright star $\sim 20$ arcsec from the {\it CXO} position, optical/IR
campaigns could neither observe the optical afterglow nor obtain
spectroscopy of the host galaxy \citep{b04,bfk+04,cb04,dct+04}.  As a
result, a spectroscopic redshift is not available for XRF\,040812.

\subsubsection{{\it HST} Observations}
\label{sec:hst_040812}

{\it HST/ACS} imaging for XRF\,040812 was carried out on 2004
September 13.4 and October 4.8 UT ($t\approx 32.1$ and 53.6 days after
the burst).  Each epoch had a total exposure time of 2000 and 2120 sec
in the F625W and F775W filters, respectively.  Observations were taken
with orientation angles chosen to minimize contamination from
diffraction spikes and saturated columns resulting from the bright
foreground star.  Using the {\it Chandra} source list provided by
\citet{pkm+04a}, we identified five unconfused, unsaturated sources in
common between the {\it Chandra} observations and our $I$-band Las
Campanas Observatory (LCO) 40-inch observations \citep{bfk+04} and
used these to tie the X-ray afterglow position to ground-based images.
The LCO and {\it HST} images were then registered, resulting in a
final positional uncertainty of $0.91$ arcsec ($2\sigma$).

Through examination of the {\it HST} images, we find an extended
source $\sim 0.33$ arcsec ($< 1\sigma$) from the nominal {\it CXO}
position and interpret it as the host galaxy of XRF\,040812.  We find
the galaxy to be relatively bright, F625W$\approx 24.50\pm 0.06$ mag,
corresponding to $R\approx 20.68\pm 0.05$ mag after correcting for the
large foreground extinction.  Comparison with the set of GRB host
galaxy magnitudes compiled by \citet{b+05} suggests that XRF\,040812
is at a relatively low redshift, $z\lesssim 0.5$.  At this redshift,
the host galaxy extinction is constrained to be $A_{V,\rm
host}\lesssim 6.1$, (90\% confidence) based on our independent
analysis of the lowest energy X-ray afterglow emission.

\begin{figure}
\vspace{-2cm}
\plotone{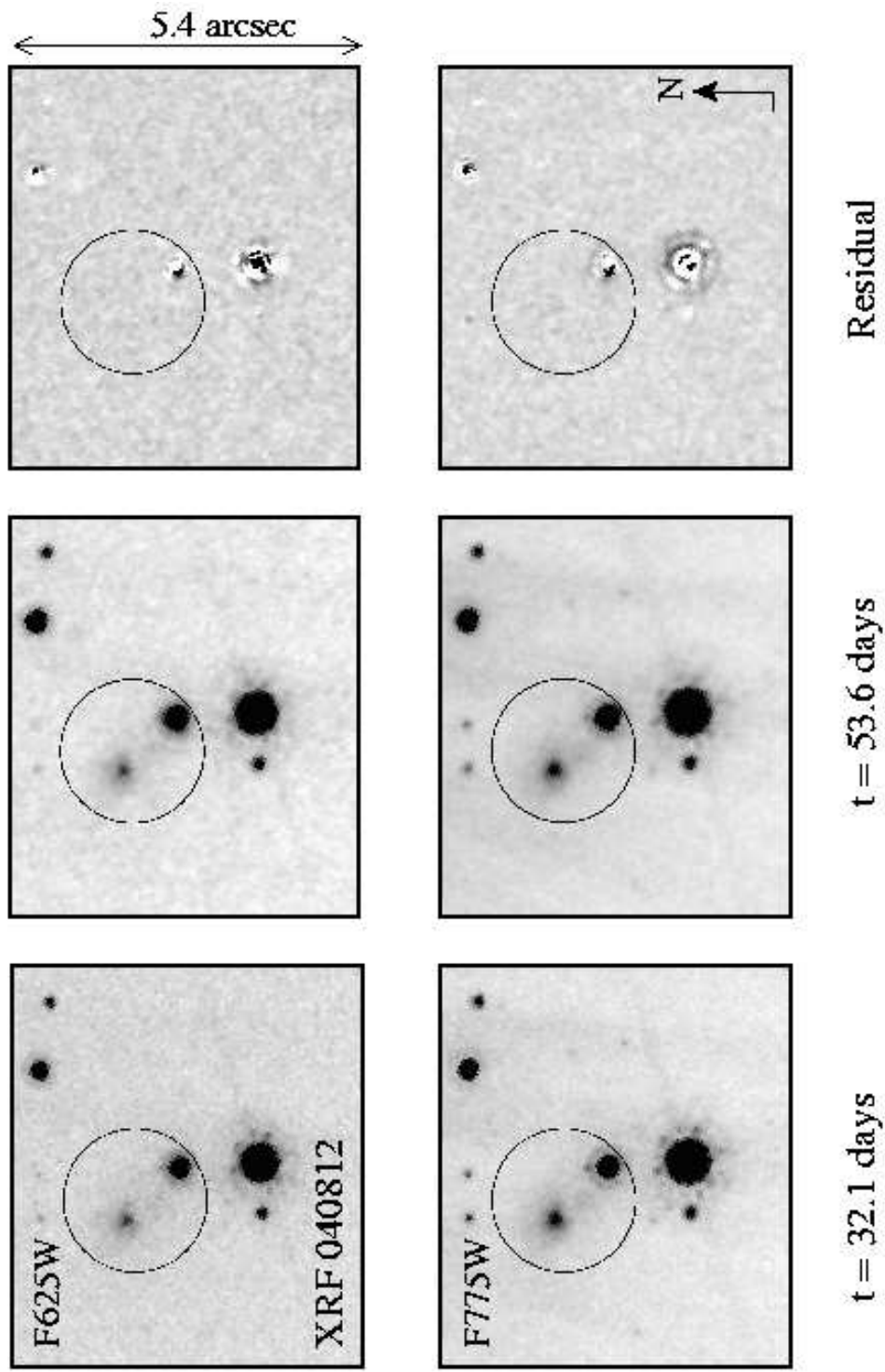}
\vspace{0cm}
\caption{Six-panel frame showing {\it HST/ACS} imaging for XRF\,040812
  at $t\sim 32.1$ days (Epoch 1) and $t\sim 53.6$ days (Epoch 2) in
  the F625W and F775W filters.  Within the X-ray afterglow position
  ($2\sigma$; circle) a host galaxy is clearly detected. By
  subtracting Epoch 2 images from Epoch 1, we produced the residual
  images, above. We apply the same stretch to all frames.  As shown in
  the residual images, there is no evidence of an optical transient
  within the host galaxy down to our {\it HST/ACS} detection limits.
\label{fig:XRF040812_HST}}
\end{figure}

Image subtraction reveals no transient sources that could be
attributed to an optical afterglow or associated supernova within the
{\it CXO} localization circle (Figure~\ref{fig:XRF040812_HST}).  We
find $3\sigma$ detection limits on the residual image of F625W $ >
27.0$ and F775W $ > 26.4$ mag ($R> 26.8$ and $I> 25.9$ mag).  Due to
the large Galactic extinction, however, the true limits are
significantly shallower, $R> 23.2$ and $I> 23.3$ mag.

\begin{figure}
\vspace{-1.5cm}
\plotone{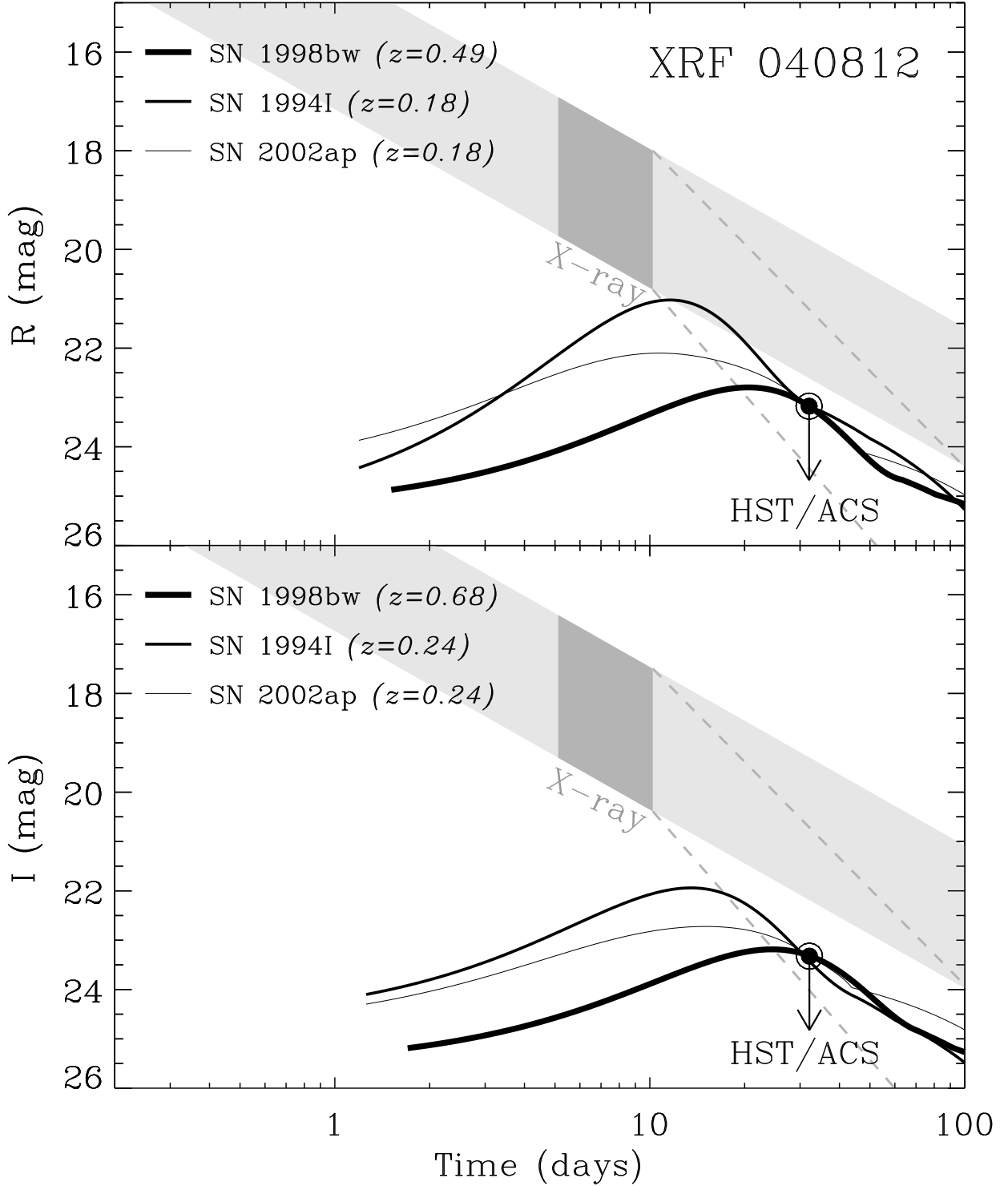}
\vspace{-1.5cm}
\caption{Constraints on a supernova associated with XRF\,040812 are
  shown for the $R$- and $I$-bands.  By extrapolating the observed
  X-ray afterglow to the optical bands, we predict a range of
  magnitudes for the optical afterglow on these same timescales (dark
  shaded polygon).  Assuming the temporal decay can be extrapolated
  outside the X-ray observation window produces the light shaded
  bands.  Dashed lines represent the case where a jet break occurs at
  the second {\it Chandra} epoch.  The marked arrows show the
  (Galactic extinction-corrected) {\it HST} constraints derived from
  image subtraction techniques described in
  Section~\ref{sec:hst_040812}.  Synthesized light-curves for SNe
  1998bw (thick), 1994I (medium) and 2002ap (thin), are shown each at
  the redshift limit to which they could be detected within our
  residual {\it HST} observations (encricled dots).
\label{fig:XRF040812_SN_curve}}
\end{figure}

Figure~\ref{fig:XRF040812_SN_curve} displays the Galactic
extinction-corrected {\it HST} limits along with the predicted optical
afterglow extrapolated from the X-ray flux in a manner similar to that
for XRF\,040701 (\S\ref{sec:hst_040701}).  Given the observed X-ray
decay, we extrapolate with $\beta\approx -0.96$ for $p\approx 2.92$ and
$\beta\approx -1.29$ for $p\approx 2.59$.  As evident from the figure, the
{\it HST} $I$-band limit is 1.9 magnitudes fainter than the
predicted optical afterglow, assuming the flux continued evolving as
$F_{\rm opt} \propto t^{-1.4}$ to the {\it HST} epoch.  If, instead, a
jet break occurred at the second {\it Chandra} epoch, the predicted
optical afterglow could be consistent with the {\it HST} non-detection.

\subsubsection{Supernova Limits}
\label{sec:sn_040812}

Since the redshift of XRF\,040812 is not known, we over-plot
synthesized light-curves for SNe 1998bw, 1994I and 2002ap at the
appropriate redshift such that the SN curves match the residual image
{\it HST} detection limit.  Supernovae placed above these redshift
limits would not be detected.  Due to the heavy foreground extinction
toward XRF\,040812, the $I$-band limits provide deeper constraints on
an associated supernova.  As shown in
Figure~\ref{fig:XRF040812_SN_curve}, a SN\,1998bw-like supernova is
ruled out for $z\lesssim 0.90$ while SN\,1994I- and SN\,2002ap-like
SNe are ruled out for $z\lesssim 0.34$ and $z\lesssim 0.35$,
respectively.  To be consistent with the estimated low-$z$ inferred
from the host galaxy brightness, an associated SN must be
significantly fainter than SN\,1998bw or suppressed due to host galaxy
extinction.

\subsection{XRF\,040916}
\label{sec:xrf040916}

\subsubsection{Prompt Emission and Afterglow Properties}
\label{sec:ag_040916}

On 2004 September 16.002 UT the {\it HETE-2} satellite discovered
XRF\,040916.  Preliminary spectral analysis revealed a dearth of
photons at $\gtrsim 10$ keV suggestive that the event was an X-ray
flash \citep{yra+04a}.  The initial localization error region was 18
arcmin in radius centered at $\alpha=23^{\rm h} 01^{\rm m} 44^{\rm
s}$, $\delta=-05^{\circ} 37' 43''$ (J2000).  A refined error box with
half the original size (545 square arcmin) was released later
\citep{yra+04b}.  Using SuprimeCam mounted on the Suburu 8.2m
telescope, \citet{kkt+04a} discovered a faint optical afterglow at
$\alpha=23^{\rm h} 00^{\rm m} 55^{\rm s}.1$, $\delta=-05^{\circ} 38'
43''$ (J2000) with a magnitude of $Rc\approx 22.3\pm 0.2$ at $t\approx
0.23$ days .  The afterglow subsequently decayed as $F_{\rm opt}
\propto t^{-1.0}$.  In comparison to GRB optical afterglows compiled
by \citet{fps+03b} and \citet{bfk+05}, XRF\,040916 is the faintest
optical afterglow ever detected on this timescale.  We note that no
spectroscopic redshift has been reported for this event.

\subsubsection{{\it HST} Observations}
\label{sec:hst_040916}

XRF\,040916 was imaged with {\it HST/ACS} on 2004 October 18.38 and
November 30.30 UT ($t\approx 32.4$ and 75.3 days after the burst) for
a total exposure time of 1930 (2058) sec in Epoch 1 and 1928 (2056)
sec in Epoch 2 in the F625W (F775W) filter.  For astrometry, we used
early-time ($t\sim 0.16$ day) $R$- and $I$-band data from the Palomar
Robotic 60-inch telescope ({\it P60}; Cenko et al., in prep) in which
the optical afterglow is clearly detected.  Twelve stars in common
between the Palomar and {\it HST} images provided an astrometric
uncertainty of $0.29$ arcsec ($2\sigma$).

\begin{figure}
\vspace{-2cm}
\plotone{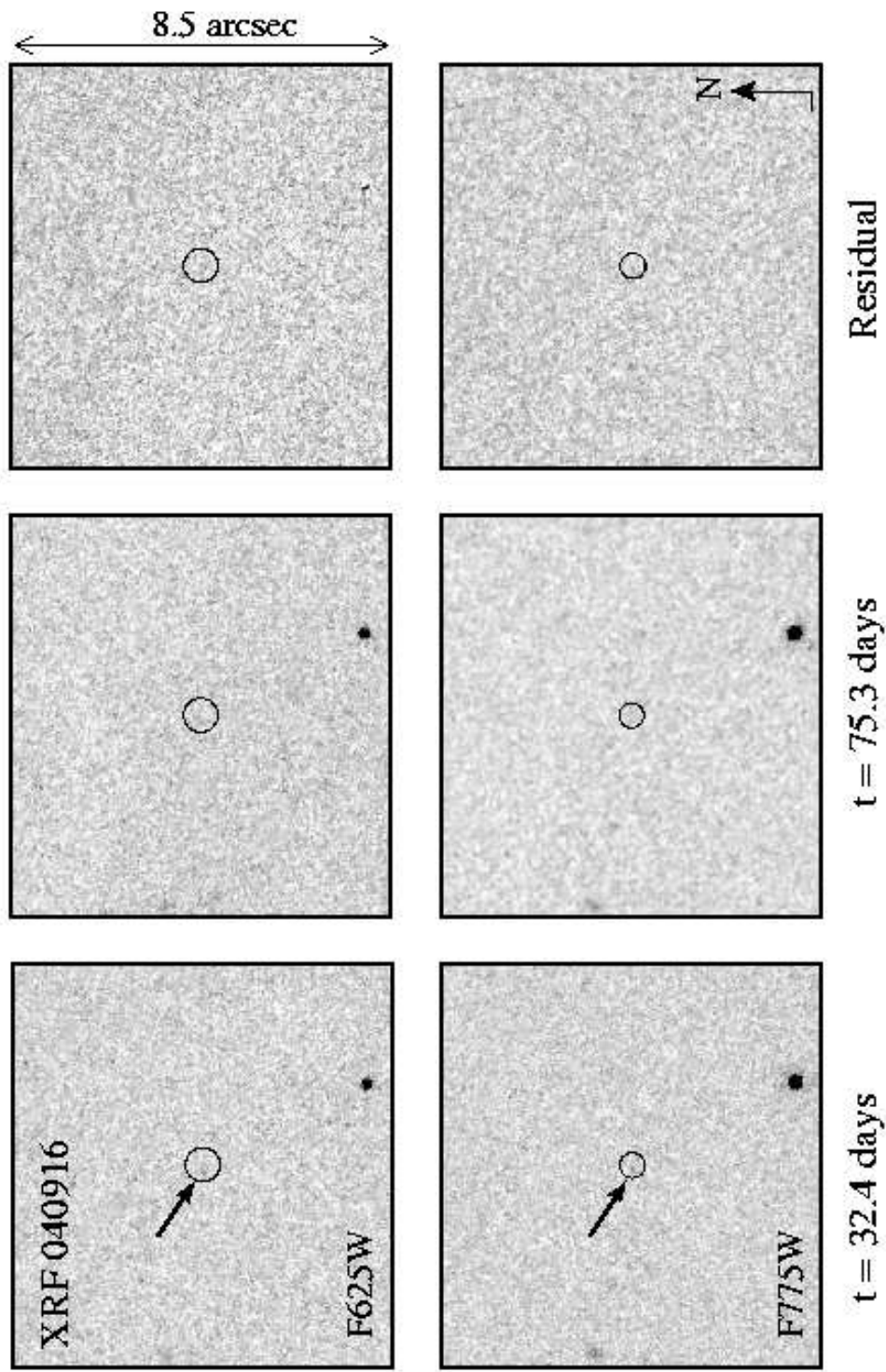}
\vspace{0cm}
\caption{Six-panel frame showing {\it HST/ACS} imaging for XRF\,040916
  at $t\sim 32.4$ days (Epoch 1) and $t\sim 75.3$ days in the F625W
  and F775W filters.  By subtracting Epoch 2 images from Epoch 1, we
  produced the residual images, above.  We apply the same stretch to
  all frames. As shown in the residual images, a faint source is
  detected in Epoch 1, coincident with the astrometrically derived
  optical afterglow position ($2\sigma$; circle).
\label{fig:XRF040916_HST}}
\end{figure}

Within the afterglow position error circle, we find a faint source
near the detection limit of our first epoch F625W and F775W images at
$\alpha=23^{\rm h} 00^{\rm m} 55.141^{\rm s}$, $\delta=-05^{\circ} 38'
42''.70$ (J2000).  Figure~\ref{fig:XRF040916_HST} shows the source is
too faint to be recovered in the residual images, implying it is just
below our 3$\sigma$ detection threshold of F625 $ > 27.9$
mag and F775W $ > 27.2$ mag ($R > 27.7$ and
$I > 26.7$ mag).  Correcting for Galactic extinction ($A_R=0.15$;
\citealt{sfd+98}), the true limits are $R > 27.5$ and $I >
26.6$ mag.

\begin{figure}
\vspace{-1.5cm}
\plotone{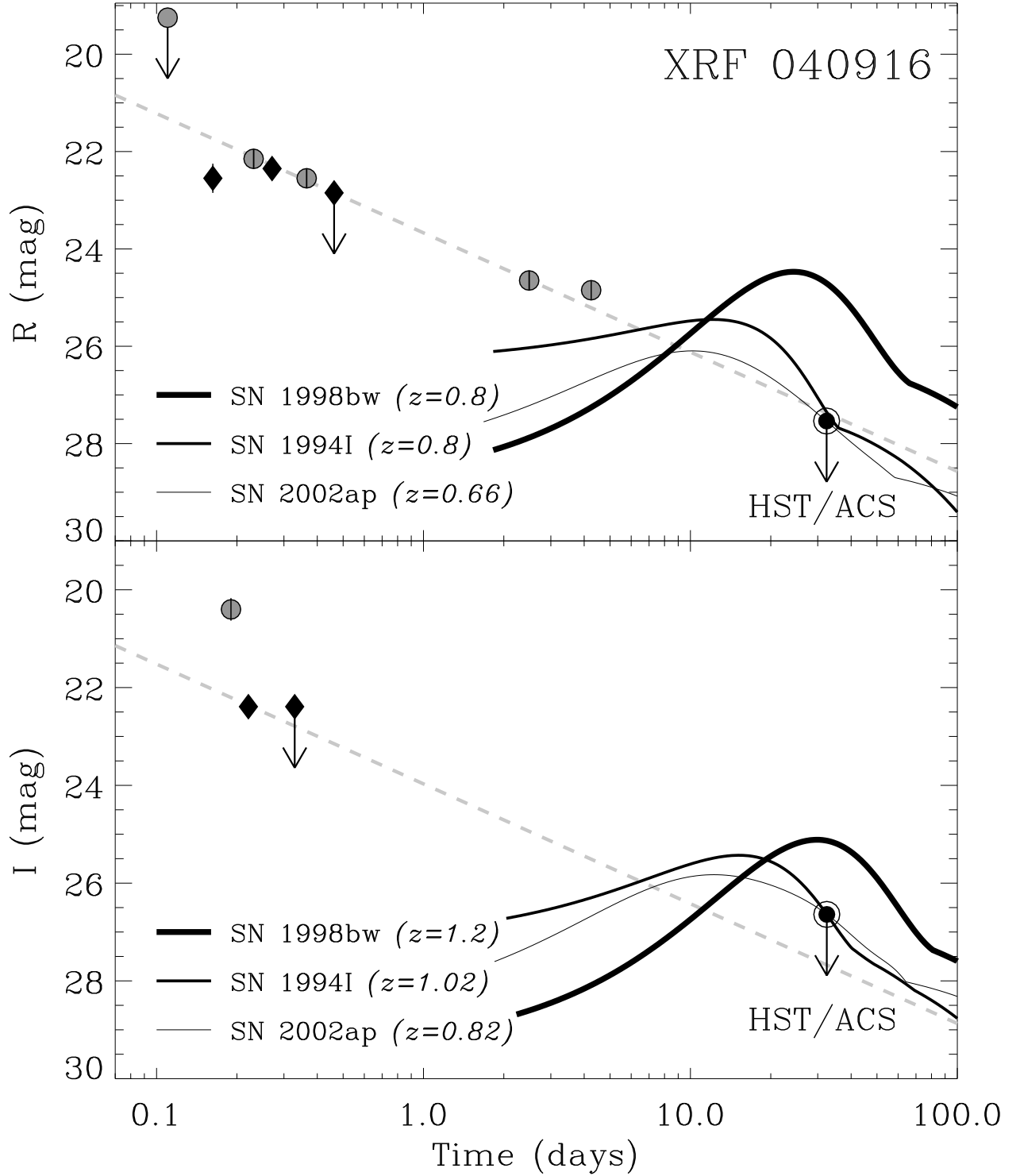}
\vspace{-2cm}
\caption{Constraints on a supernova associated with XRF\,040916.
  Extinction corrected measurement for the optical afterglow include
  data from the robotic Palomar 60-inch telescope (diamonds) and from
  the GCNs (grey circles) and are plotted for the $R$- (upper panel) and
  $I$-band (lower panel). The temporal decay of the $R$-band afterglow
  is well fit by $t^{-1.0}$ (grey dashed line) which we scale to fit
  the $I$-band Palomar 60-inch data as well.  The observed magnitude
  of the faint {\it HST} transient (encircled dot) is consistent with
  the extrapolated OT decay. Synthesized light-curves for SNe 1994I
  (medium line) and 2002ap (thin line) are shown at the redshift for
  which the SN brightness matches that of the residual source.  At
  $z=1.2$ (the redshift limit for our $I$-band light-curve synthesis)
  SN\,1998bw (thick line) would still be $\gtrsim 3$ magnitudes
  brighter than the {it HST} residual source (encircled dot).  We note
  this non-detection is consistent with a higher redshift, as possibly
  suggested by the faintness of the optical afterglow and host galaxy
  (\S\ref{sec:sn_040916}).
\label{fig:XRF040916_SN_curve}}
\end{figure}

In Figure~\ref{fig:XRF040916_SN_curve} we show the {\it HST} limits
along with early-time data from {\it P60} and from the GCNs
\citep{kkt+04a,kas+04,h04}, all corrected for Galactic extinction. The
early data are well fit with a $F_{\rm opt}\propto t^{-1.0}$ decay.
We note that the early detection by \citet{h04} is inconsistent with
our $I$-band data.  The {\it HST} limits are consistent with the
extrapolated optical afterglow, assuming the flux decayed without
steepening until the epoch of {\it HST} observations.

\bigskip

\subsubsection{Supernova Limits}
\label{sec:sn_040916}

Since the redshift of XRF\,040916 is not known, we over-plot
synthesized light-curves for SNe 1994I and 2002ap each at the
appropriate redshift such that the SN curves match the residual image
{\it HST} detection limit.  Supernovae with light-curves similar to
SNe 1994I and 2002ap would not be detected above $z\approx 1.02$ and
$z\approx 0.82$, respectively.  Synthesized light-curves for
SN\,1998bw at $z=0.8$ and $z=1.2$ represent the limits to which we can
confidently extrapolate the rest-frame SN spectrum. These light-curves are
$\sim 3$ and $\sim 1.5$ magnitudes brighter than the observed {\it HST}
limits in the $R$- and $I$- bands, and therefore suggest that either
XRF\,040916 is at higher redshift (e.g. $z > 1.2$) or it is associated
with a lower luminosity SN.

We note that XRF\,040916 is the only event within our {\it HST} survey
for which we do not detect the host galaxy.  Given our {\it HST}
detection limit and the small foreground extinction, this implies the
host galaxy is fainter than $R\approx 27.5$ mag.  We compare this
limit with faint GRB host galaxies; {\it all} of the GRB hosts with $R
> 26$ mag are at redshift $z > 1.4$ \citep{b+05}.  Assuming the GRB
host galaxy luminosity function also applies to XRFs, this implies
that XRF\,040916 is at a similarly high redshift, far beyond the
$z\sim 1.2$ redshift limit out to which we can detect an associated
SN.

\section{A Summary of XRF-SN Searches}
\label{sec:summary}

We now present a global summary of all SN searches in XRFs to date,
including {\it HST} campaigns for XRFs 011030, 020427, 020903, 040701,
040812, 040916 and a deep ground-based effort for XRF 030723.  In
compiling these results, we also discuss the available constraints on
host galaxy extinction, which could suppress any emission from an
associated SN.

The only XRF for which we find evidence suggestive of an accompanying
supernova is XRF\,020903 ($z=0.25$).  Based on the observed flattening of the
optical light-curves at $t\sim 30$ days, and the identification of
broad features in the optical spectrum, we propose that XRF\,020903
was associated with a supernova up to $\sim 1$ magnitude fainter than
SN\,1998bw at maximum light.  

In the case of XRF\,040701, our foreground extinction-corrected {\it
HST} detection limit is $\sim 6$ mag fainter than SN\,1998bw at this
redshift. Our analysis of the X-ray afterglow spectra reveals that the
rest-frame host galaxy extinction is constrained to $A_{V,\rm host}
\lesssim 2.8$, implying a conservative upper limit on the brightness
of the associated SN to be $\sim 3.2$ mag fainter than SN\,1998bw and
fainter than all of the GRB-SN studied to date.  Taken together, XRFs
020903 and 040701 (the only two XRFs with redshifts in our sample)
imply that at least some of the XRF-associated SNe are considerably
fainter than SN\,1998bw.

Due to the heavy foreground extinction toward XRF\,040812, our ability
to detect a SN\,1998bw-like event would only be possible for
$z\lesssim 0.7$.  In this case, analysis of the X-ray afterglow does
not provide a strong constraint on the host galaxy extinction, since
the lowest energy photons are absorbed by the Galaxy.  Our limits are
more constraining for the case of XRF\,040916, where we would be
sensitive to a SN\,1998bw-like supernova beyond $z\sim 1.2$.  The lack
of a SN detection, however, is consistent with a high redshift as
possibly suggested by the faintness of the host galaxy and optical
afterglow.

Deeper constraints have previously been reported based on {\it
  HST/STIS} data for two additional XRFs without known redshifts, XRFs
  011030 and 020427 \citep{lpk+04}.  Thanks to the broad throughput of
  the STIS Clear filter, the sensitivity extends red-ward of $I$-band,
  thereby enabling SN detection to $z\approx 1.5$ before UV
  blanketing suppresses the observed emission.  \citet{lpk+04} showed
  that a SN\,1998bw-like supernova would be detectable to $z\sim 1.5$
  for both XRFs 011030 and 020427.  Using the X-ray afterglow data for
  these two bursts, \citet{lpk+04} estimated their host galaxy
  extinction to be $A_V \lesssim 1.7$ and $A_V \lesssim 2.5$,
  respectively, assuming a moderate redshift of $z\sim 0.5$.  At lower
  redshift, the host galaxy extinction required to suppress a
  SN\,1998bw-like supernova is inconsistent with the X-ray limits,
  suggesting that these two bursts are either located at higher
  redshift or associated with low luminosity SNe.  We note that a
  firm redshift limit of $z\lesssim 2.3$ has been reported for
  XRF\,020427 based on the lack of Ly$\alpha$ absorption down to 3800
  \AA\ within the optical spectrum \citep{vb03}.

While there were no {\it HST} observations taken for XRF\,030723, we
utilize the deep ground-based afterglow observations reported by
\citet{fsh+04} to constrain the emission from an underlying supernova.
Despite claims that the optical rebrightening at $t\sim 10$ days is
due to an associated supernova \citep{tdm+04,fsh+04}, we conclude that
the observed optical/NIR variability is dominated by afterglow
emission since neither the $R-K$ color nor the optical to X-ray
spectral index vary on this timescale (Fox et al., in prep.;
\citealt{bss+04}).  We therefore adopt the $R$-band afterglow
light-curve as an effective upper limit on the flux of an accompanying
SN.  We derive the most robust constraint from an observation at
$t=24.78$ days with $R=25.08\pm 0.09$.  This limit is sufficiently
deep to rule out a SN\,1998bw-like supernova at $z\lesssim 0.8$.  We
note that while the redshift of XRF\,030723 is not known, the lack
of Ly$\alpha$ absorption in optical spectra limits the redshift to
$z\lesssim 2.3$ \citep{fsh+04}.

\begin{figure}
\plotone{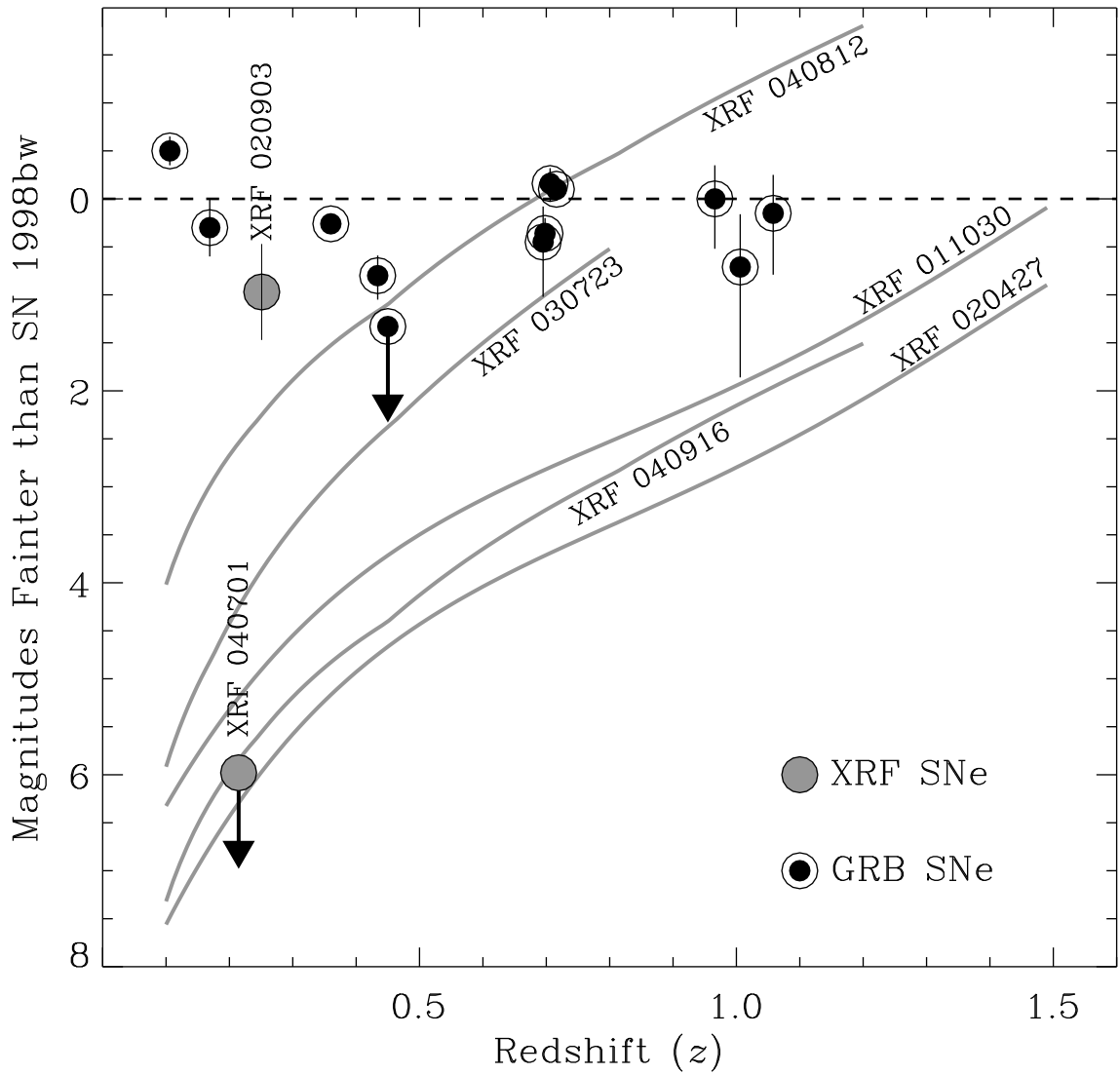}
\caption{A compilation of constraints for SN\,1998bw-like supernovae
  associated with X-ray flashes.  For each XRF we adopt the (Galactic
  extinction-corrected) {\it HST} limit for the brightness of an
  associated supernova.  At a given redshift, the difference between
  the observed {\it HST} limit and the synthesized SN\,1998bw
  light-curve represents how faint the supernova must be (with respect
  to SN\,1998bw) in order to go undetected in our images.  We plot
  this magnitude difference against redshift for the seven XRFs with
  deep, late-time observations.  Results for XRFs 011030 and 020427
  have been taken from \citet{lpk+04}.  For XRF\,040701 we show the
  limit including negligible host galaxy extinction, but note that this could
  be as high as $A_{V,\rm host}\sim 3$ magnitudes.  For XRF\,020903 we
  plot the value based on late-time {\it HST} observations but note
  that the peak magnitude could be considerably brighter (see
  \S\ref{sec:sn_020903} and Fig.~\ref{fig:XRF020903_SN_curve}).  Also
  shown are the confirmed GRB-SN detections compiled from the
  literature (see \S\ref{sec:summary}).
\label{fig:SN1998bw_delta_mag}}
\end{figure}

We summarize the limits on XRF-associated supernovae for these seven
XRFs in Figure~\ref{fig:SN1998bw_delta_mag}.  Also shown are the peak
SN magnitudes (relative to SN\,1998bw) for limits (GRB\,010921,
\citealt{pks+03}) and secure detections of GRB-associated supernovae
compiled from the literature (GRBs 970228, 980703, 990712, 991208,
000911, 011121, 020405, 021211 \citealt{zkh04}; GRB\,030329,
\citealt{log+04,dtm+05}; GRB\,031203, \citealt{mtc+04}; GRB\,041006,
\citealt{sgn+05}).  This figure highlights the spread of luminosities
implied for the supernovae associated with high energy cosmic
explosions.  The GRB-SNe clearly show a spread in their
peak brightness.  Based on XRFs 020903 and 040701, XRF-SNe appear to
show a similar spread.  Assuming a modest redshift of $z\sim 1$ for
the other five XRF-SNe implies an even larger spread and might
suggest than XRFs are associated with systematically fainter SNe.

\section{Discussion}
\label{sec:conclusions}

We presented results from our {\it HST/ACS} search for the supernovae
associated with the XRFs 020903, 040701, 040812, 040916 and extended
this sample by including published results for SN searches in XRFs
011030, 020427 and 030723.  We find strong evidence (photometric and
spectroscopic) for a SN\,1998bw-like supernova (dimmed by $\lesssim 1$
mag) in association with XRF\,020903 ($z=0.25$).  This finding
conclusively associates XRFs with the death of massive stars for the
first time.  In the case of XRF\,040701 ($z=0.21$), our {\it HST}
limit is $\sim 6$ magnitudes fainter than SN\,1998bw which cannot be
accounted for by host galaxy extinction.  Based on these two events
(XRFs 020903 and 040701), we conclude that at least some
XRF-associated SNe exist but can be significantly fainter than
SN\,1998bw.

\begin{figure}
\plotone{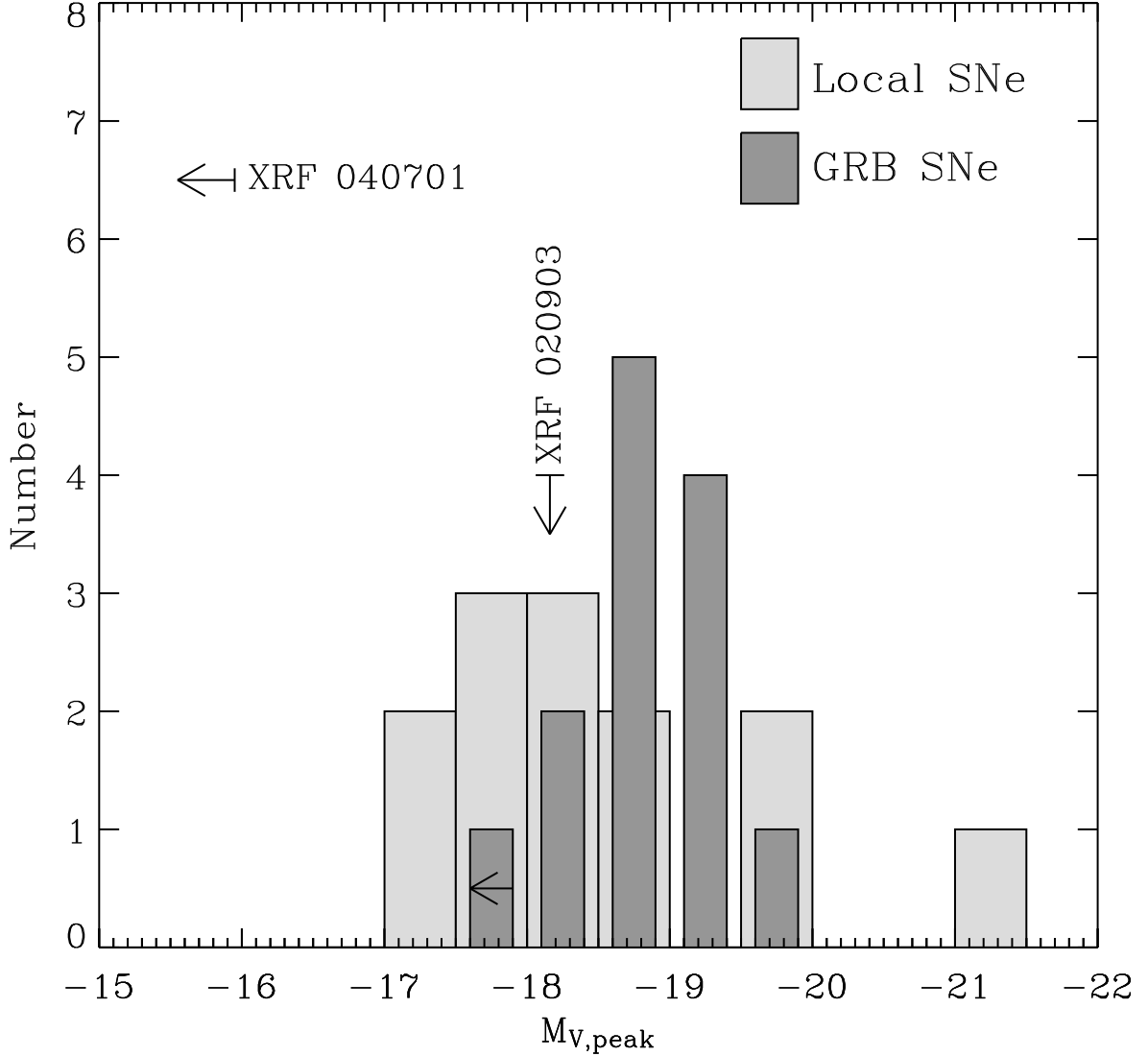}
\caption{Histogram of peak optical magnitudes (rest frame $M_V$) for
  local Type Ibc supernovae (light grey) and GRB associated SNe (dark
  grey).  Locations of the SNe associated with XRFs 020903 and 040701
  (with $A_{V,\rm host}=2.8$ mag) are marked with arrows.  There is a
  significant spread in peak brightness demonstrated by the local SNe
  and the GRB-SNe samples.  The SN associated with XRF\,020903 is
  consistent with this observed diversity. GRB-SNe have been compiled
  from the literature (see \S\ref{sec:summary}).  Local SNe include
  SN\,1983N \citep{cwb+96}, SN\,1983V \citep{cwp+97}, SN\,1984L
  \citep{t87}, SN\,1987M \citep{fps90}, SN\,1990B \citep{csp+01},
  SN\,1991D \citep{bbt+02}, SN\,1992ar \citep{cps+00}, SN\,1994I
  \citep{rvh+96}, SN\,1997ef \citep{inn+00}, SN\,1999as
  \citep{hbn+01}, SN\,1999ex \citep{shs+02}, SN\,2002ap
  \citep{fps+03a}, and SN\,2003L (Soderberg {\it et al.}, in prep).
\label{fig:SN_hist}}
\end{figure}

In Figure~\ref{fig:SN_hist} we compile peak magnitudes for local Type
Ibc supernovae and GRB-associated SNe.  While the GRB-SN population
tend to lie at the bright end of the local SN luminosity distribution
there is a pronounced $\sim 2$ magnitude spread in their observed peak
brightness which produces significant overlap with the local SNe.
We emphasize that the SN associated with XRF\,020903 is consistent
with this observed spread, however, a supernova associated with
XRF\,040701 would be considered faint even in the context of the local
SN population.

Since peak luminosity correlates roughly with the mass of $^{56}$Ni
synthesized in the explosion, these observations imply a significant
spread in the Nickel yield from GRB and XRF explosions.  Most
importantly, as revealed from low luminosity GRB/XRF SNe,
engine-driven relativistic explosions do not necessarily produce more
$^{56}$Ni than local SNe Ibc, which are not associated with
relativistic ejecta \citep{bkf+03}.  This result is bracketed by two
extremes: events with a large Nickel output can be associated with
weak engine-driven explosions harboring only a small amount of
relativistic ejecta (e.g. SN\,1998bw/GRB\,980425), and those with a
low Nickel yield can be produced in strong engine-driven explosions
characterized by copious amounts of energy coupled to highly
relativistic jetted material.  In conclusion, we find evidence that
Nickel production and engine activity represent independent parameters
of the GRB/XRF explosion mechanism, each of which can be individually
tuned.

\begin{acknowledgements}
The authors are grateful for support under the Space Telescope Science
Institute grant HST-GO-10135.  A.M.S. acknowledges support by the NASA
Graduate Research Fellowship Program Research. E.B. is supported by
NASA through Hubble Fellowship grant HST-HF-01171.01 awarded by the
STScI, which is operated by the Association of Universities for
Research in Astronomy, Inc., for NASA, under contract NAS 5-26555.
A.G. acknowledges support by NASA through Hubble Fellowship grant
\#HST-HF-01158.01-A awarded by STScI.  D.C.L. is supported by an NSF
Astronomy and Astrophysics Postdoctoral Fellowship under award
AST-0401479.

\end{acknowledgements}

\clearpage

\bibliographystyle{apj1b} 
%\bibliography{journals_apj,xrf_refs}

\clearpage

\begin{deluxetable}{clrcrccc}
\tablecaption{{\it HST} Observation Log}
\tablewidth{0pt} \tablehead{ 
\colhead{Target} & \colhead{Date Obs} & \colhead{$\Delta t$} & 
\colhead{Exp. Time} & \colhead{Filter} & \colhead{{\it HST} mag\tablenotemark{a}} & 
\colhead{Extinction\tablenotemark{b}} & \colhead{Johnson mag\tablenotemark{c}} \\
\colhead{} & \colhead{(UT)} & \colhead{(days)} & 
\colhead{(sec)} & \colhead{} & \colhead{(AB)} & 
\colhead{$A_{\lambda}$} & \colhead{(Vega)} \\ } 

\startdata
XRF\,020903 & 2002 December 3.8 & 91.4 & 1840 & F606W & $24.53\pm 0.03$ & $A_R=0.09$ & $R=24.23\pm 0.03$ \\
\nodata & 2003 June 30.6 & 300.2 & 1920 & F606W & \nodata & \nodata & \nodata \\
XRF\,040701 & 2004 August 9.6 & 39.1 & 1820 & F625W & $> 27.81$ & $A_R=0.13$ & $R > 27.48$ \\
\nodata & 2004 August 30.5 & 60.0 & 1820 & F625W & \nodata & \nodata & \nodata \\
\nodata & 2004 August 9.7 & 39.1 & 1940 & F775W & $> 26.83$ & $A_I=0.09$ & $I > 26.30$ \\
\nodata & 2004 August 30.5 & 60.0 & 1940 & F775W & \nodata & \nodata & \nodata \\
XRF\,040812 & 2004 September 13.3 & 32.0 & 2000 & F625W & $> 26.99$ & $A_R=3.61$ & $R > 23.18$ \\
\nodata & 2004 October 4.8 & 53.5 & 2000 & F625W & \nodata & \nodata & \nodata \\
\nodata  & 2004 September 13.4 & 32.1 & 2120 & F775W & $> 26.38$ & $A_I=2.62$ & $I > 23.32$ \\
\nodata & 2004 October 4.8 & 53.5 & 2120 & F775W & \nodata & \nodata & \nodata \\
XRF\,040916 & 2004 October 18.4 & 32.4 & 1930 & F625W & $> 27.89$ & $A_R=0.15$ &  $R > 27.54$ \\
\nodata & 2004 November 30.3 & 75.3 & 1928 & F625W & \nodata & \nodata & \nodata \\
\nodata & 2004 October 18.4 & 32.4 & 2058 & F775W & $> 27.19$ & $A_I=0.64$ & $I > 26.64$ \\
\nodata & 2004 November 30.4 & 75.4 & 2056 & F775W & \nodata & \nodata & \nodata \\
\enddata
\tablenotetext{a}{From residual image.  Observed magnitude, not corrected for foreground extinction.}
\tablenotetext{b}{Galactic extinction from \citet{sfd+98}.}
\tablenotetext{c}{From residual image.  Corrected for foreground extinction using $A_{\lambda}$ given in column 7.}
\label{tab:hst}
\end{deluxetable}

\clearpage

\clearpage

\clearpage

\clearpage

\clearpage

\clearpage

\clearpage

\clearpage

\clearpage

\clearpage

\clearpage

\end{document}